# Study of electron emission from 1D nanomaterials under super high field Nonlinear Strong-field

Chi Li[1†], Ke Chen[1†], Mengxue Guan[2], Xiaowei Wang[3], Xu Zhou[4], Feng Zhai[5], Jiayu Dai[3], Zhenjun Li[1], Zhipei Sun[6,7], Sheng Meng[2*], Kaihui Liu[4*] Qing Dai[1*]

[1]Division of Nanophotonics, CAS Center for Excellence in Nanoscience, National Center for Nanoscience and Technology, Beijing 100190, China.

[2]Beijing National Laboratory for Condensed Matter Physics and Institute of Physics, Chinese Academy of Science, Beijing, 100190, China

[3]Department of Physics, College of Science, National University of Defense Technology, Changsha 410073, China

[4]School of Physics, Academy for Advanced Interdisciplinary Studies, Collaborative Innovation Center of Quantum Matter, Peking University, Beijing 100871, China.

[5]Department of Physics, Zhejiang Normal University, Jinhua 321004, China.

[6]Department of Electronics and Nanoengineering, Aalto University, Tietotie 3, FI-02150 Finland

[7]QTF Centre of Excellence, Department of Applied Physics, Aalto University, FI-00076 Aalto, Finland

*Correspondence: daiq@nanoctr.cn (Q.D.), khliu@pku.edu.cn (K.L.), smeng@iphy.ac.cn (S.M.).
†These authors contributed equally to this work.

**Photoemission driven by a strong electric field of near-infrared or visible light, referred to as strong-field photoemission, produces attosecond ($10^{-18}$ s) electron pulses that are synchronized to the waveform of the incident light[1-4], and this principle lies at the heart of current attosecond technologies[5,6]. However, full access to strong-field photoemission regimes at near-infrared wavelengths based on solid-state materials is restricted by space-charge screening and material damage at high optical-field strengths[7,8], which significantly hampers the realization of predicted attosecond technologies, such as ultra-sensitive optical phase modulation[9,10]. Here, we demonstrate a new type of strong-field photoemission behaviour with extreme nonlinearity -- photoemission current scales follow a 40$^{th}$ power law of the optical-field strength, making use of sub-nanometric carbon nanotubes and 800 nm pulses. As a result, the total photoemission current depends on the carrier-envelope phase with a greatly improved photoemission current modulation depth of up to 100%, which has not previously been achieved[7]. Time-dependent density functional calculations reveal the completely new behaviour of the optical-field induced tunnelling emission process[11,12] directly from the valence band of the carbon nanotubes, which is an indication of full access to a strong-field photoemission regime. Furthermore, the nonlinear dynamics are observed to be tunable by changing the binding energy of the valence-band-maximum, as confirmed by Simpleman model calculations. We believe that such extreme nonlinear photoemission from nanotips offers a new means of producing extreme temporal-spatial resolved electron pulses. These results additionally provide a new design philosophy for attosecond electronics and optics by making use of tunable band structures in nanomaterials.**

Strong-field photoemission arises when the optical-field is sufficiently strong to vibrate the vacuum barrier at the material surface, and then a purely perturbative description of the emission process in the photon picture is no longer sufficient[13]. In such a regime, electrons are liberated in a fraction of an optical-cycle[14-16]. By employing near-infrared or visible laser pulses, it is possible to generate electron pluses with an attosecond temporal resolution and with a high degree of synchronization to the incident optical waveform[17]. Not only does this advance time-resolved electron characterization into an attosecond time domain, but it also provides attosecond control and measurement methodology[5]. Strong-field photoemission is thus a fundamental part of attosecond technologies such as attosecond electron microscopy[18], petahertz electronic devices[19-21], attosecond light sources[22,23], optical-phase detectors[6,24], among others.

Over the past decade, strong-field photoemission has been achieved with nanostructures at wavelengths down to 800 nm[25,26]. However, it has been proved to be the so-called strong-field above-threshold-photoemission (SF-ATP) regime[2,27], which still involves multiphoton process. The strong optical-field induces the shift of multiphoton photoelectron peaks (channel-closing effect)[27,28] and a plateau in the high-energy part of the electron spectrum[2]. In such regimes, the photoemission has a relatively low nonlinear response[7,27,29-31] -- the photoemission current ($I$) normally scales with a low order ($N \approx 2$) power law of incident optical-field strength ($F$). This has led to a far less sensitive carrier-envelope-phase (CEP) effect than that predicted by previous theoretical approaches[10]. At a stronger optical-field, strong-field photoemission may transit into an optical-field-emission (OFE) regime. In this case, the optical-field is strong enough to create a penetrable barrier, and tunnelling takes place from states within the vicinity of the Fermi level, in which a much stronger nonlinear response is expected, because $I$ generally increases exponentially[11] with $F$. The long-awaited strong CEP effect in solid-state materials may then be observed.

Currently, full access to OFE at near-infrared wavelengths is generally restricted by space-charge screening[7,8] and material damage[32]. Using carbon nanotubes (CNTs), however, with ultrahigh $\beta$ and ultrasharp tip[33,34], together with their high mechanical strength, it may be possible to full access OFE (Fig. 1a). A high optical-field enhancement factor ($\beta$) of the sub-nanometric CNT tips plays a key role in reducing the required $F$, which greatly reduces the risk of material damage when using a high intensity laser. Furthermore, the space-charge screening effect can largely be overcome using a much sharper nanotip with radially diverging particle trajectories and a large static (lightning rod) field enhancement for rapid charge escape[27].

We report on a new strong-field photoemission behaviour, in which the $I$-$F$ curve deviates dramatically from conventional low nonlinear scaling in an SF-ATP regime, based on a CNT cluster emitter with sub-nanometric tips (see Supplementary Methods for details). The emitter is driven by 100 fs ultrafast optical pulses centred around 800 nm (see Supplementary Methods for details of the setup used). As shown in Fig. 1b, the log-log plot of the *I-F* curve clearly indicates three different operation regimes, with two "kinks". The first "kink" has been commonly reported to mark the transition from a photon-driven (multiphoton photoemission, MPP)[35,36] to a strong-field regime (SF-ATP)[27,37,38], occurring for a Keldysh parameter[39] $1<\gamma<2$. In the present work, using the simulated value of $\beta = 20$ for CNTs (Extended Data Fig. 3b), the estimated value of $\gamma$ at the first "kink" ($\gamma_1$) is 1.1, while at the second kink ($\gamma_2$) it is 0.7. We speculate that the second "kink" marks the transition from SF-ATP to OFE. These three regimes are therefore multiphoton photoemission (blue dots), SF-ATP (green dots), and OFE (red dots). The corresponding diagram illustrating these regimes is given in Fig. 1c. The full-access OFE regime, indicated by the second "kink", is reported here for the first time for solid-state materials. In this unique OFE process, an ultra-high photoemission nonlinearity is observed, where the *I* displays an extremely high order ($N = 40$) power law of the *F*.

To investigate such extreme nonlinear photoemission behaviour, we undertook carrier-envelope phase (CEP) dependent photoemission measurements[7,29]. Ultrashort laser pulses of around 3 optical cycles (7 fs pulse duration) with an 800 nm central wavelength were used (see Supplementary Methods for setup). The CEP-stabilized measurement of an *I-F* curve is shown in Fig. 2a. In CEP-stabilized measurements, the total photoemission current is greatly reduced compared with the 100 fs case (Fig. 1b), due to a greatly reduced pulse width. Therefore, only the photoemission regime for the high *F* region in the *I-F* curve is clearly observed, in which *I* is above the noise current level ($10^{-11}$ A). In the high-field part (with corresponding $\gamma < 0.66$), a similarly high nonlinearity ($N = 40$) is observed.

We fixed the laser intensity (peak optical-field ~1.3 V/nm), and controlled the CEP ($\Delta\phi_{CEP}$) for photoemission measurements, as presented in Fig. 2b. As expected, in the extreme nonlinear OFE process, *I* is modulated effectively by changing the CEP[7]. By tuning CEP for 6 $\pi$, the measured data points can be fitted to a Cosine curve quite well, providing solid evidence of a strong-field photoemission mechanism[2]. The modulation depth reaches up to 100 %, which is around twice as high as previously reported values obtained using metallic tips[7]. This result clearly shows that full access to OFE can offer more sensitive control of the photoemission process with CEP, which shows promise for further improvement of the temporal precision of attosecond measurements and control[2].

To depict the physics of electron dynamics of OFE as distinct from SF-ATP regimes, we calculate the photoemission process using the framework of time-dependent density functional theory (TDDFT), which covers both the photon (weak) and field (strong) regimes in a single description[40,41] (see Supplementary methods). Calculations are performed for a zigzag (10, 0) single-walled CNT with a diameter of ~0.85 nm, which is within the range of tube diameters used in the experiments (see Extended Data Fig. 3). An incident Gaussian light pulse with an 800 nm central wavelength and a full width at half maximum of 7 fs is used in our calculations. The calculated *I-F* curve is shown in Fig. 3a, which shows the same three linear behaviours as the experimental results (Fig. 1b).

The excitation behaviours in field-driven tunnelling and photon-driven regimes are intrinsically different. In the photon-driven regime, the energy difference between the initial and final states of an electron transition is equal to the integer multiple of the incident photon energy[42], and the momentum should be conserved. This results in discrete and selective excitation energy levels that are largely dependent on the excitation photon energies[43]. In the field-driven tunnelling emission regime, the tunnelling probability of electrons is far less dependent on the photon energy, rather it mainly relies on both the field strength and the initial energy level. When the vacuum barrier is bent downwards by a strong optical-field, a triangular barrier forms at the emitter/vacuum interface, and the electron tunnelling probability decreases exponentially with the energy level (barrier thickness increasing). The detailed excitation behaviours are simulated by tracking the changes of electrons in the states near the Fermi Level with different *F* (see Supplementary Methods), as shown in Figs. 3b, c, both of which show in the right-hand panels the density of states (DOS) of the CNT model for positioning. With a low *F* (1 V/nm, Fig. 3b), the peaks of the lost electrons are mainly located at two energy levels (0 eV and -1.5 eV). Simultaneously, the peaks of gained electrons are mainly located at three energy levels (0 and 1.4 eV). The energy intervals of these peaks are almost equal to the incident photon energy (1.55 eV), which is clearly a multiphoton (i.e., the photon-driven) regime. By increasing $F$ to 5 V/nm (Fig. 3c), the excitation behaviour changes, becoming fundamentally different from the photon-driven regime. Most of the electrons are excited from states at the valence band maximum (VBM, around 0.9 eV). Furthermore, the number of excited electrons exhibits exponential decay when the energy levels go deeper, which is a phenomenon typical of field-driven tunnelling emission. A similar calculation result is also obtained for a 400 nm wavelength for the CNT model (Extended Data Fig. 4). These results fully confirm that the present observed highly nonlinear photoemission regime belongs to the OFE from the VBM of semiconductors.

The nonlinearity of OFE can be tuned by engineering the band structure of the emitting materials. According to the Wentzel-Kramers-Brillouin approximation, the relationship between the electron tunnelling probability and the local field strength displays an exponential function (see Supplementary Methods), of which the exponent is largely dependent on its binding energy (BE). Therefore, the nonlinearity of OFE can be largely tuned by changing the BE of the VBM. To investigate this behaviour, an over-threshold emission (OTE) process is applied to change the BE of the VBM of the emitter. In the present setup, the Fermi level of all CNTs in the cluster is fixed at the same level as their connected substrate -- a heavily doped silicon chip. The band structure of the whole system is schematically illustrated in Fig. 4a. When illuminated with high intensity laser pulses to reach a saturation emission current (over-threshold), the CNT tips are gradually burnt away from a lower to a higher BE. Thus, the BE of the entire CNT emitter is reduced after the OTE process. By carefully controlling such processes, the $I$-$F$ curves at different stages may be obtained, as shown in Fig. 4b. It is clear that the nonlinearity increases with increasing BE.

The trend can be further understood by computing the photoemission current using the modified two-step Simpleman model[11], which considers all potential emissions from the occupied states near the Fermi level (see Supplementary Methods). The calculated $I$-$F$ curves of the CNT results with different BEs (0.16 eV, 0.24 eV, 0.49 eV, see Extended Data Fig. 5 for the density of states with different BEs) are exhibited in Fig. 4c. As expected, the nonlinearity increases with the BE, while *I* decreases as a tradeoff. Although *I* can be compensated by an enhanced *F*, it is limited by the finite mechanical strength of the CNT. Consequently, the photoemission nonlinearity, referred to as the modulation efficiency, must be balanced against the photoemission current for future applications.

In conclusion, a record nonlinearity of photoemission with a power factor of up to 40 is achieved for a CNT emitter, leading to a highly efficient (up to 100%) CEP modulation. TDDFT calculation results show that the extreme nonlinearity is a result of optical-field driven electron tunnelling from the VBM of the CNT, an indication of full access to OFE. Furthermore, such nonlinearity can be efficiently tuned by engineering the band structures of the emitter, which may be realized by controlling the chirality and doping levels of the CNTs. The scheme is believed to be universal for other nanomaterials with tunable band structures. The extreme nonlinear OFE may further improve the temporal precision of ultrafast electron pulses and optical-phase measurement down to attosecond timescales. We predict that such highly efficient optical-field control of the electron motion in nanostructures is a new form of quantum electronics, and may pave the way for the generation, measurement, and application of attosecond electronics and optics.

Figures and Captions

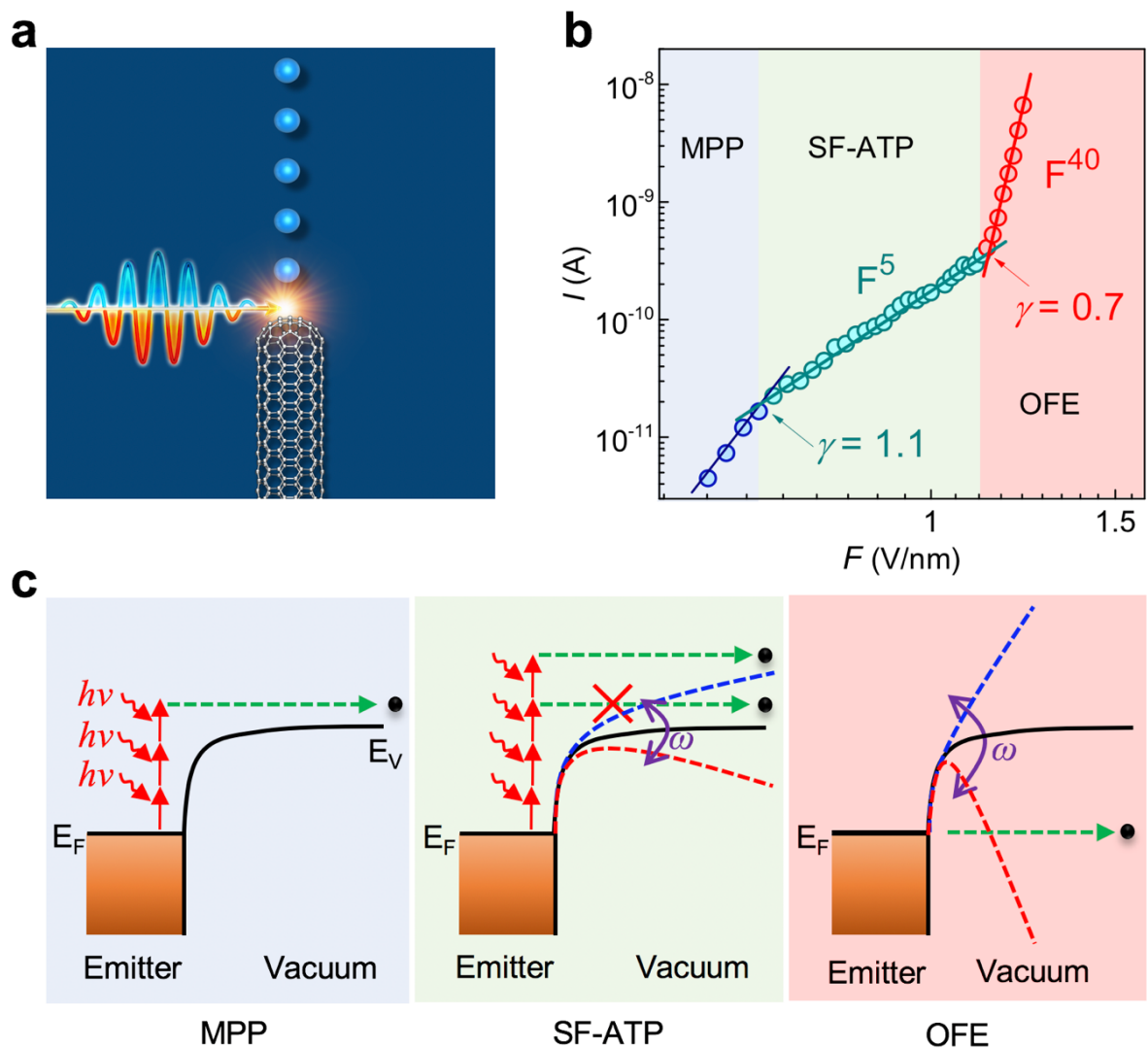

**Figure 1 | Operation Principle of OFE. a**, Electron (balls) are emitted from a CNT emitter, driven by negative half-cycles (red part) of a laser pulse. **b**, Experimentally obtained log-log plot of the *I-F* curve, driven by 100 fs laser pulses with a central wavelength of 820 nm. Three regimes are shown by different colours: blue, multi-photon photoemission, MPP; green, SF-ATP, red, OFE. A 40$^{th}$ power law is observed in the OFE regime. **c**, Corresponding mechanisms of these three photoemission regimes: Left-hand panel, MPP. A electron absorbs several photons (*hv*) to overcome the vacuum level ($E_V$). Middle panel, SF-ATP. In the positive cycle, $E_V$ is elevated (blue dashed line), which induces a channel closing (red cross) for low-order channels. Thus, electrons must absorb more photons to overcome the elevated $E_V$. Right-hand panel, OFE. Optical-field is strong enough to induce a tunnelling emission from states near the Fermi level ($E_F$). *ω,* angular frequency of the incident optical-field. Black balls, electrons. Green dashed arrow line, photoemission channels.

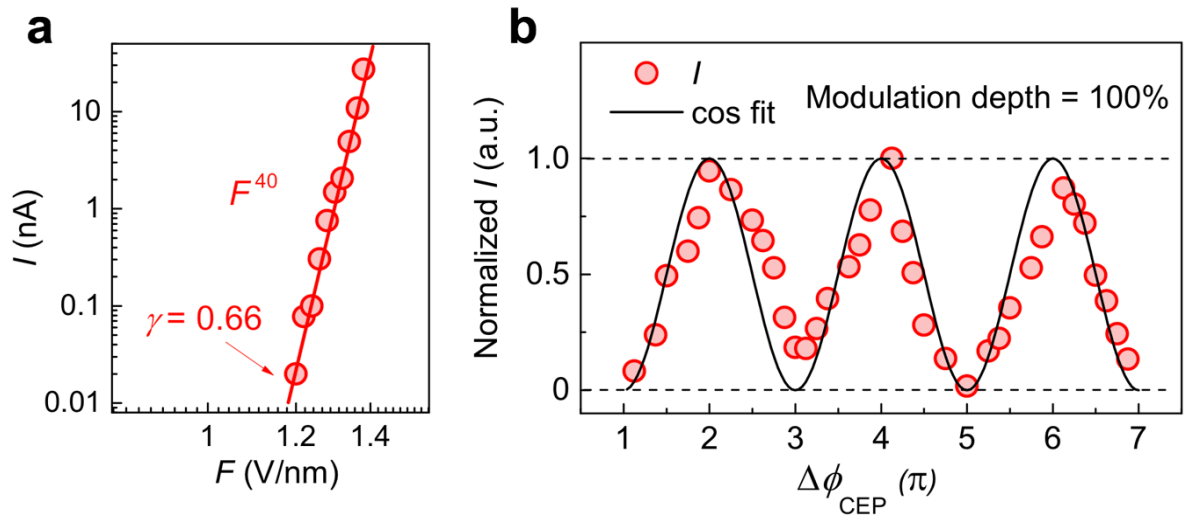

**Figure 2 | CEP-dependent photoemission current.** Experimental results obtained at 7 fs with few-cycle laser pulses centred around 800 nm. **a,** CEP stabilized measurement of the *I-F* curve. The calculated $\gamma = 0.66$ suggests that the OFE process is accessed. A power order of 40 is obtained. **b,** The CEP-dependent photoemission current at a fixed laser intensity with a peak optical-field strength of 1.3 V/nm. The current is normalized by dividing all current data by the maximum current value, in order to see the modulation depth clearly, while the calculation of the modulation depth is unaffected.

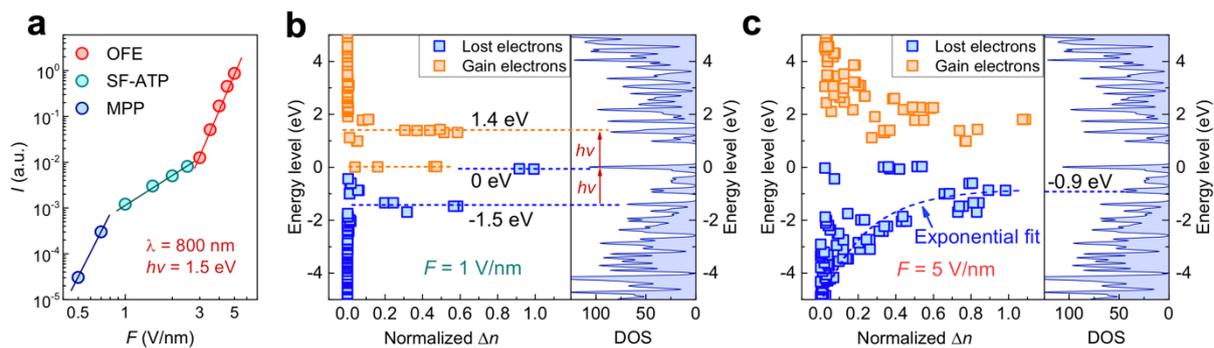

**Figure 3 | TDDFT calculation results. a,** The calculated *I-F* curve of a CNT model (see Supplementary Methods) reveals three regimes, including OFE, SF-ATP, and MPP regimes, consistent with the experimental data. The simulated laser pulse is centred around 800 nm (*hv* = 1.55 eV) with a pulse width of 7 fs. **b,** Excitation states at *F* = 1 V/nm. The normalized number (Δ*n*) of lost (blue squares) and gained (orange squares) electrons at different energy levels clearly shows peaks with an energy interval approaching *hv*, which demonstrates photon-driven behaviour. **b,** Excitation states at *F* = 5 V/nm, Δ*n* of the lost electrons decreases exponentially as the energy level goes deeper, which demonstrates field-driven tunnelling behaviour. The right-hand panels of (B) and (C) show the density of states (DOS) of the model.

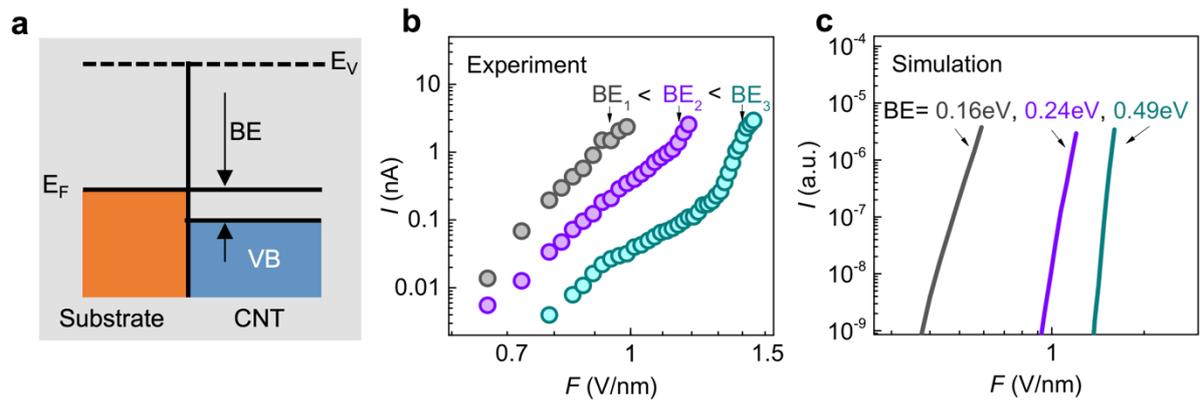

**Figure 4 | Band-structure dependent photoemission nonlinearity. a,** Schematic band-structure of a CNT cluster emitter. $E_V$, vacuum level; $E_F$, Fermi level; VB, valence band. **b,** Experimentally obtained *I-F* curves of a CNT cluster emitter before and after the OTE process. Before OTE, $BE_1$; After the first round of OTE, $BE_2$; After the second round of OTE, $BE_3$. Thus, $BE_1<BE_2<BE_3$. Clearly, highly nonlinear OFE is more prominent in CNTs with higher BE. **c,** Simulated *I-F* curves of three CNT models with different BEs. Higher BE is associated with greater nonlinearity.

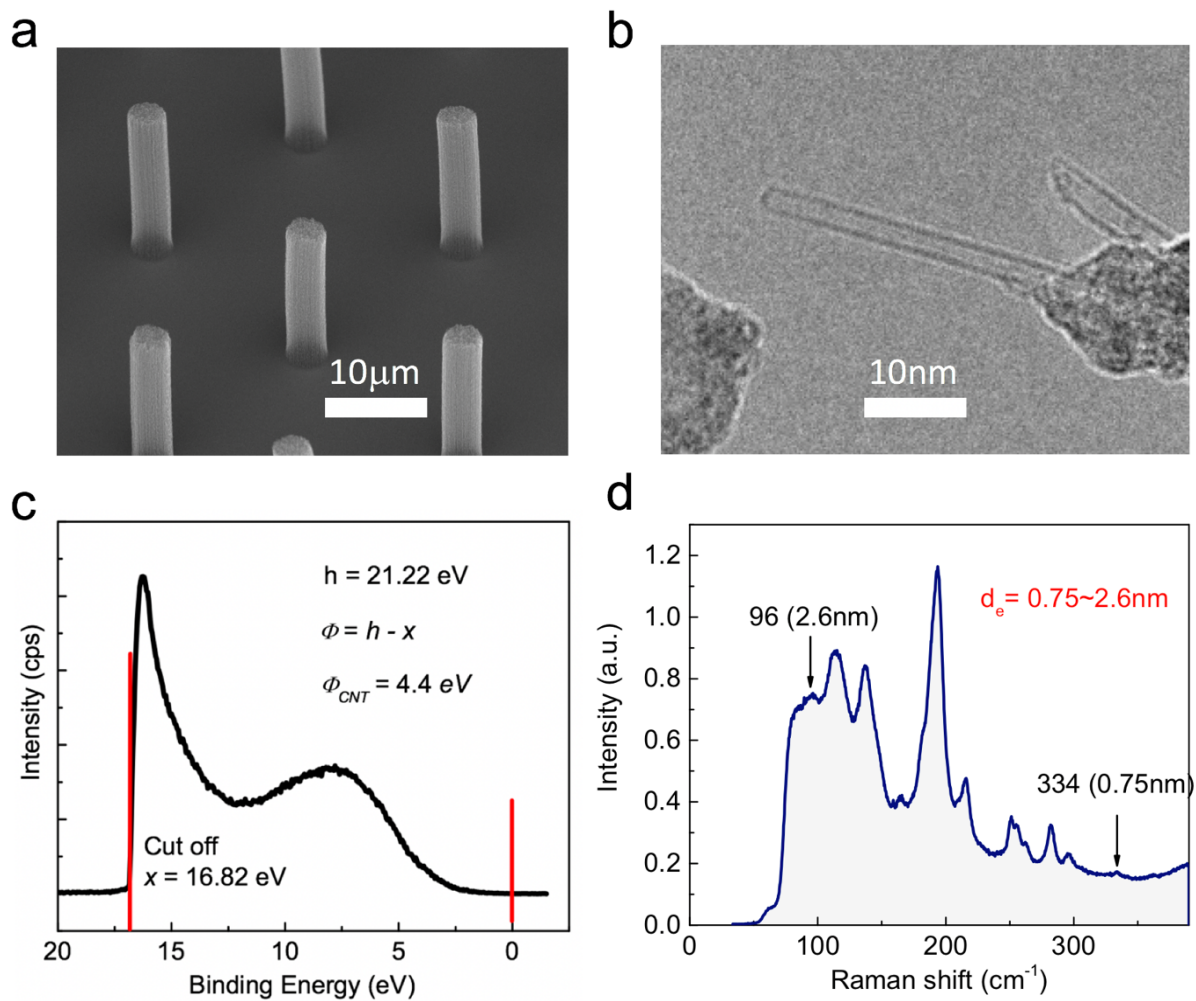

**Extended Data Figure 1 | Characterization of CNTs. a,** Scanning electron microscopy image of the as-grown CNT cluster array. **b,** High resolution transmission electron microscopy image of an individual single-walled carbon nanotube. **c,** Ultraviolet Photoelectron Spectroscopy of CNT, showing a work function of ~4.4 eV. **d,** Raman spectrum of the CNT cluster, indicating diameters ranging from 0.75 to 2.6 nm.

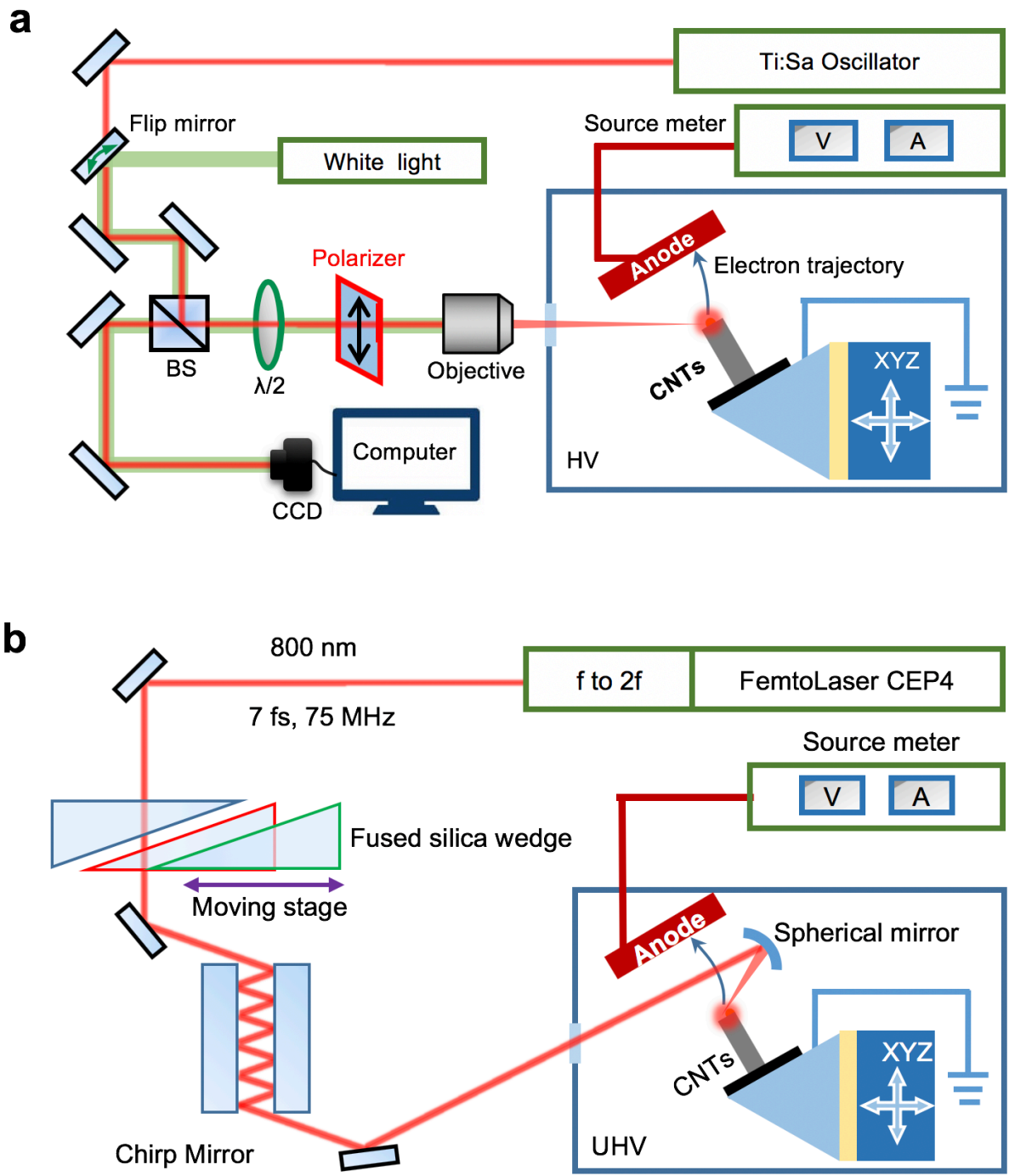

**Extended Data Figure 2 | Experimental setups. a**, 100 fs laser system and measurement of optical-field dependent photoemission current. **b**, 7 fs few-cycle laser system and measurement of CEP-dependent current.

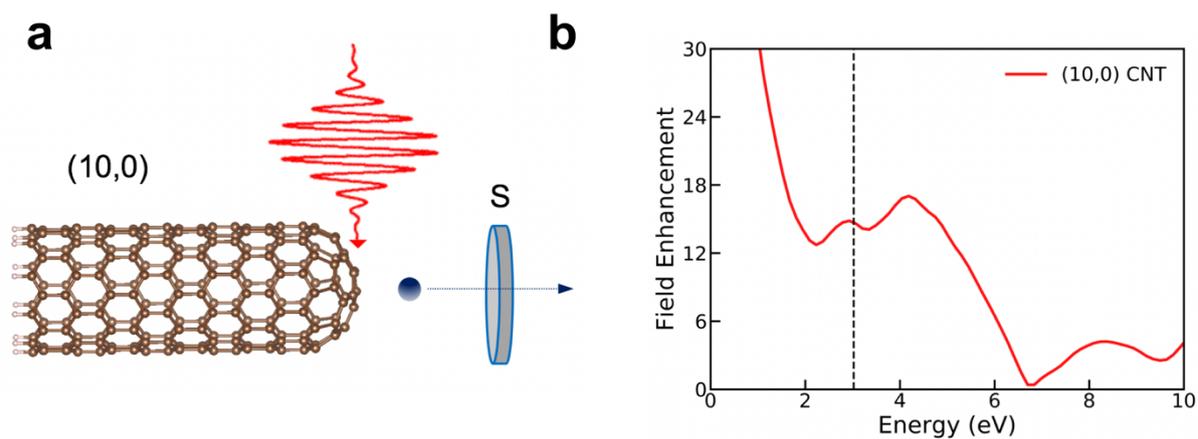

**Extended Data Figure 3 | Simulation model**. **a,** A capped CNT (10, 0) tip. Due to the absence of periodic boundary conditions in the molecular calculations, it is necessary to saturate the carbon dangling bonds with hydrogen atoms, yielding a $C_{200}H_{10}$ tube. The polarization of the incident optical-field is consistent with the axial tube. An electron detection plane is set in front of the capped end at a distance of 0.2 nm, this can count the electrons that cross the plane in both vertical directions, and can detect their kinetic energy. **b,** Calculated field enhancement along the axis of (10, 0) CNT as a function of photon energy.

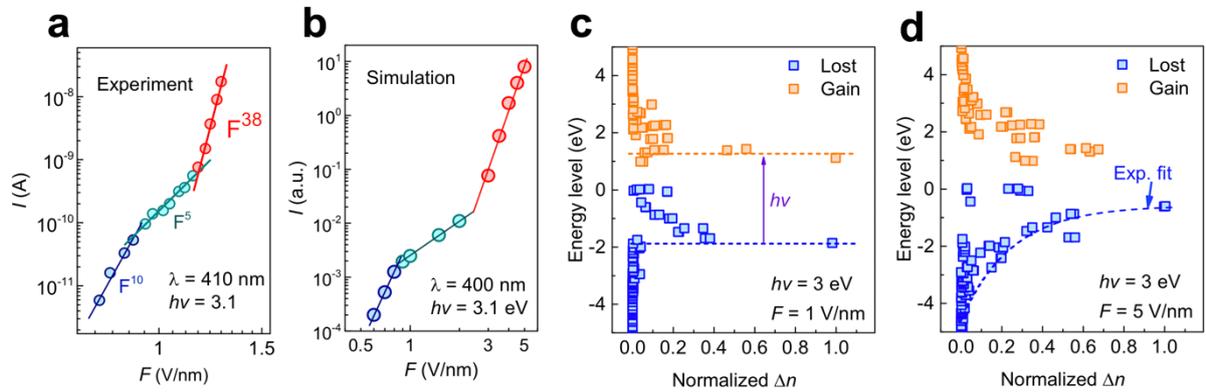

**Extended Data Figure 4 | TDDFT results of photoemission current and excitation states.**
**a,** Experimentally obtained I-F curve. **b,** The simulated laser pulses are centred around 400 nm ($h\nu$ = 3.1 eV) with a pulse width of 7 fs. **c,** Excitation states at $F$ = 1 V/nm. The normalized number ($\Delta n$) of lost (blue squares) and gained (orange squares) electrons for different energy levels clearly shows peaks with an energy interval approaching $h\nu$, which demonstrates photon-driven behaviour. **d,** Excitation states at $F$ = 5 V/nm, $\Delta n$ of the lost electrons decreases exponentially with energy level, which demonstrates field-driven tunnelling behaviour.

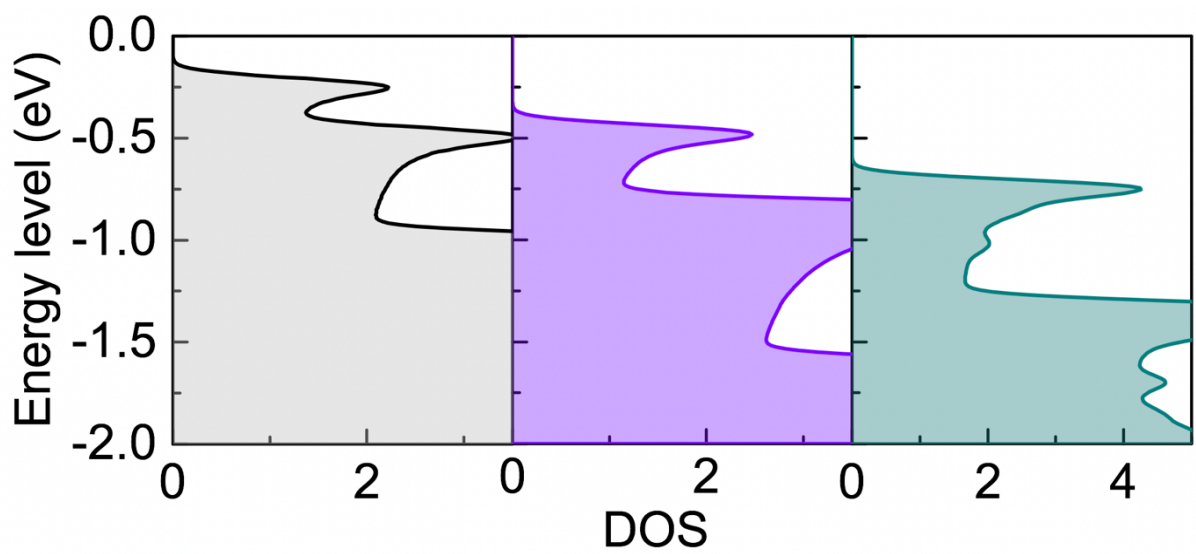

**Extended Data Figure 5 | DOS of three CNT models with different BEs.**